\documentclass{moriond}

\def\Journal#1#2#3#4{{#1} {\bf #2}, #3 (#4)}

\def\NIMA{{\em Nucl. Instrum. Methods} A}

\def\PRL{\em Phys. Rev. Lett.}
\def\PRD{{\em Phys. Rev.} D}

\def\mco{\multicolumn}

\def\be{\begin{equation}}
\def\ee{\end{equation}}
\def\bea{\begin{eqnarray}}
\def\eea{\end{eqnarray}}

\newcommand{\Photo}{\includegraphics[height=35mm]{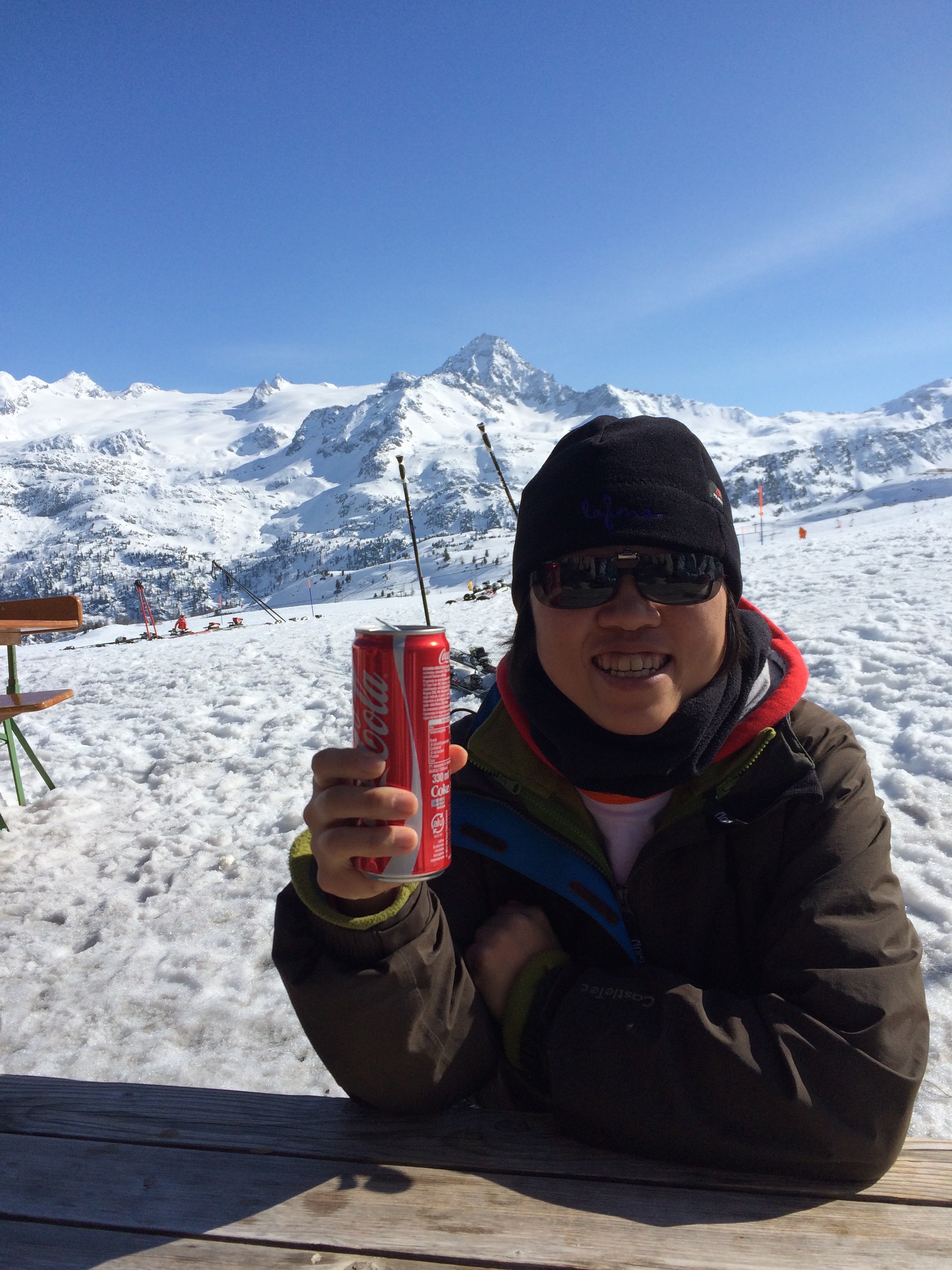}}
\usepackage{eqnarray,amsmath}
\usepackage{cite}

\begin{document}
\vspace*{4cm}
\title{Recent Results from Daya Bay Reactor Neutrino Experiment}

\author{ B. Z. HU\\
on behalf of the Daya Bay collaboration }

\address{Department of Physics, National Taiwan University,\\
No. 1, Sec. 4, Roosevelt Rd., Taipei 10617, Taiwan}

\maketitle\abstracts{
The Daya Bay reactor neutrino experiment announced the discovery of a non-zero value of $\sin^22\theta_{13}$ with significance better than 5$\sigma$ in 2012. The experiment is continuing to improve the precision of $\sin^22\theta_{13}$ and explore other physics topics. In this talk, I will show the current oscillation and mass-squared difference results which are based on the combined analysis of the measured rates and energy spectra of antineutrino events, an independent measurement of $\theta_{13}$ using IBD events where delayed neutrons are captured on hydrogens, and a search for light sterile neutrinos.}

\section{Introduction}
The neutrino flavor eigenstates are linear combinations of the mass eigenstates, given as\\
\begin{equation}
| \nu_\alpha \rangle = \sum^3_{i=1}  U_{\alpha , i}^{\ast} | \nu_i \rangle, 
\label{eq:murnf}
\end{equation}
where $\alpha$ represents the neutrino flavors, $e$, $\mu$ and $\tau$,  $i$ represents the mass states, and $U_{\alpha i}$ is the unitary matrix known as the Pontecorvo-Maki-Nakagawa-Sakata (PMNS) mixing matrix, 
\begin{eqnarray}
U_{PMNS}  & = &  \nonumber 
  \begin{pmatrix} 
1 & 0  & 0 \\ 
0 & C_{23} & S_{23}  \\
0 & -S_{23} & C_{23} 
 \end{pmatrix}
   \begin{pmatrix} 
C_{13}   		& 0 	& e^{-i\delta}S_{13} \\ 
0  			& 1 	& 0 \\
-e^{i\delta}S_{13} &   0 & C_{13}
 \end{pmatrix}
   \begin{pmatrix} 
C_{12}   & S_{12}  & 0 \\ 
- S_{12}  & C_{12} & 0 \\
0 &   0 		          & 1
 \end{pmatrix},
 \end{eqnarray}
where $C_{ij}$ is $\cos \theta_{ij}$, $S_{ij}$ is $\sin \theta_{ij}$ and $\delta$ is the CP violating phase. In 2012, the Daya Bay collaboration has published the first non-zero results with a significance of 5.2 standard deviations by using the data from the period with six antineutrino detectors \cite{An:2012eh, An:2013uza}.

For electron antineutrinos with energy E traveling a distance L in vacuum, the survival probability is given by
\begin{equation}
P(\overline{\nu}_e \rightarrow \overline{\nu}_e)=1-\sin^22\theta_{13} \sin^2 \big( \frac{\Delta m^2_{ee}L}{4E} \big) - \cos^4 \theta_{13}\sin^22\theta_{12} \sin^2 \big( \frac{\Delta m^2_{21}L}{4E} \big),
\label{eq:Psur}
\end{equation}
where 

\begin{equation}
\sin^2 \big( \frac{\Delta m^2_{ee}L}{4E} \big) \equiv \cos^2 \theta_{12} \sin^2 \big( \frac{\Delta m^2_{31}L}{4E} \big)+\sin^2 \theta_{12} \sin^2 \big( \frac{\Delta m^2_{32}L}{4E} \big)
\end{equation}
and 

\begin{equation}
\Delta m^2_{ij} \equiv m^2_i - m^2_j.
\end{equation}

Using $ \Delta m^2_{32} = \Delta m^2_{31} =2.32 \times 10^{-3} \mathrm{eV^2}$, $\Delta m^2_{21} = 7.59 \times 10^{-5} \mathrm{eV^2}$ and $\sin^22\theta_{12} = 0.861^{+0.026}_{-0.022}$ \cite{Adamson:2011ig, Abe:2008aa}, the distance L for the first minimum of $P(\bar \nu_e \to
\bar \nu_e)$ for reactor antineutrinos is $\sim$1.6 km. The value of $\theta_{13}$ can be determined from the observed
deficit of antineutrino flux detected via the inverse beta decay.

To have a precise measurement of $\theta_{13}$, it requires an optimized baseline, high statistics, and low systematic uncertainties and low backgrounds. For the Daya Bay experiment, the far site detectors are near the location of maximum oscillation effect. The Daya Bay power plant has a thermal power of 17.4 $\mathrm GW_{th}$, which generate a very high electron antineutrino flux. The target mass for each antineutrino detector is 20 tons in Gd-loaded liquid scintillator region and 20 tons in non-loading liquid scintillator region. The reactor and detector -related systematic errors are reduced by using identically-designed detectors at the near and far sites for a far/near relative measurement. 

\section{Daya Bay Experiment}
The Daya Bay experimental site is located in the southern part of China near Shenzhen city. The Daya Bay nuclear power complex consists of six reactor cores with a total of 17.4 $\mathrm GW_{th}$ thermal power. There are three underground experimental halls (EHs): two near halls and one far hall. The near-hall detectors measure the neutrino flux from the reactor cores with negligible effect from the mixing angle $\theta_{13}$. The far-hall detectors can measure the neutrino oscillation effect due to $\theta_{13}$. Each near hall contains two antineutrino detectors (ADs) and the far hall contains four ADs \cite{An:2012side}. The data taking with 6 ADs started on Dec. 24, 2011, and was paused in the summer of 2012 to install the last two ADs. The data taking has been on-going with 8ADs since October 2012.  

\subsection{Antineutrino Detector Design }\label{subsec:fig}

Each of the eight functionally identical antineutrino detectors consists of three zones separated by acrylic vessels. The inner zone is the antineutrino target containing 20 tons of Gadolinium-loaded liquid scintillator (GdLS). The middle zone is the gamma catcher containing 20 tons of liquid scintillator (LS). The outer zone, filled with 40 tons of mineral oil (MO), shields background radiations. There are 192 8'' PMTs mounted on eight ladders in each AD. To improve the light collection, there are reflectors on the top and bottom of the outer acrylic vessel. On top of each ACU there are three automatic calibration
units (ACUs), each containing three sources, LED, $\mathrm{^{68}Ge}$ and $\mathrm{^{241}Am-^{13}C+^{60}Co}$. Calibrations are performed once a week.

\subsection{Muon Veto System}
The muon veto system consists of a water pool instrumented with PMTs as the cherenkov detectors and 4 layers of RPC tracking detectors.
The former is composed of the inner water shield (IWS) and outer water shield (OWS), to detect cosmic ray muons and to shield neutrons and gammas from rock. The latter covers the water pool to provide further muon tracking information. 

\subsection{Detector Response}
\begin{figure}[htbp]
\centering
\includegraphics[width=0.5\linewidth]{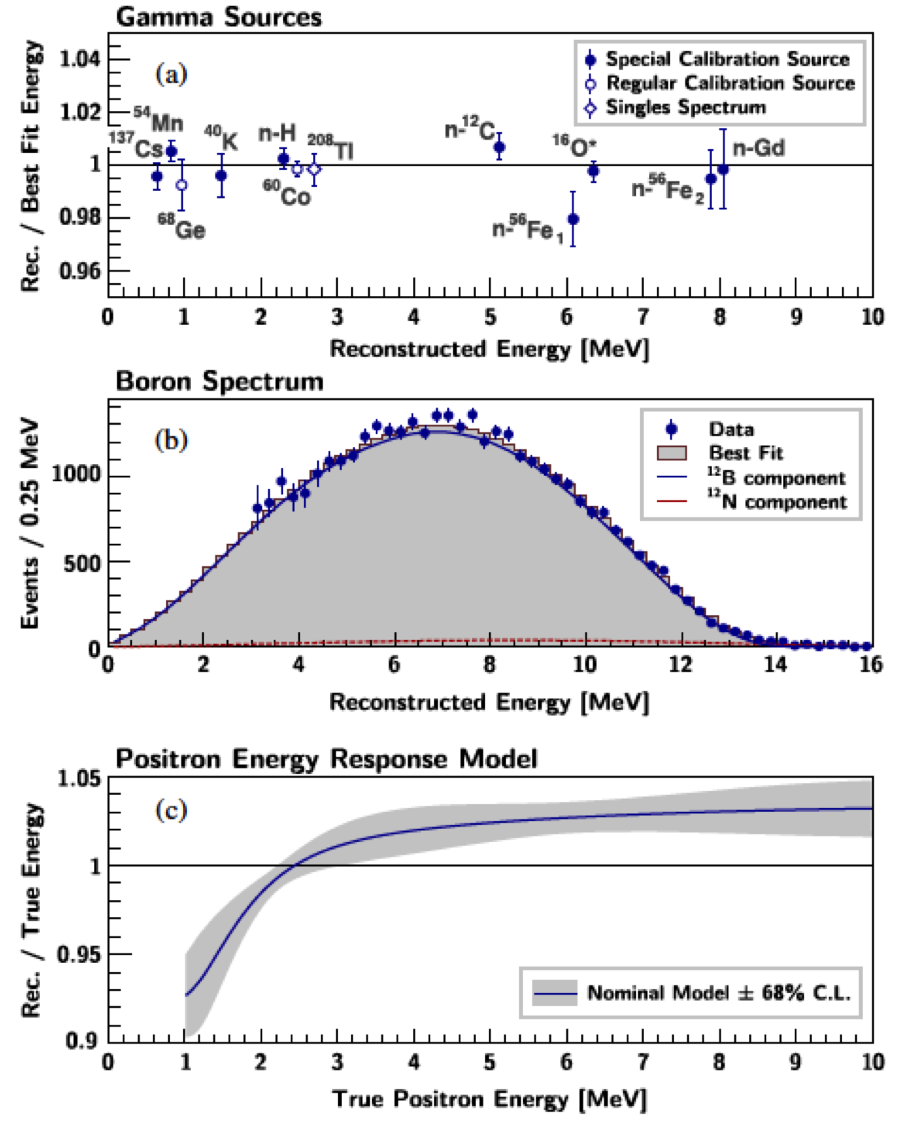}
\caption{(a) Ratio of the reconstructed to the best-fit energies of $\gamma$ lines from calibration source and the singles spectra. (b) Reconstructed energy spectrum (points) compared to the sum (shaded area) of the $\mathrm{^{12}B}$ (solid line) and $\mathrm{^{12}N}$ (dashed line) components of the best-fit energy response model. (c) AD energy response model for positrons.}
\label{fig:nonL}
\end{figure}

Many calibration sources (ACU sources, $\mathrm{^{137}Cs}, \mathrm{^{54}Mn}, \mathrm{^{40}K}$, $\mathrm{^{241}Am-^{9}Be}$ and $\mathrm{^{239}Pu-^{13}C}$) and environmental sources ($\mathrm{^{40}K}, \mathrm{^{208}Tl}$ and $n$ capture on H, C and Fe) were used to measure the energy resolution and study the nonlinearity of detector response. 
The energy resolution is $7.5/\sqrt{E_{vis}/\mathrm{MeV}}$. 
The nonlinearity of detector response, caused by the liquid scintillator and the readout electroncs characteristics, has a minimal impact on the oscillation angle measurement, but is more relevant for the measurement of the reactor antineutrino mass difference. The energy response model is obtained semi-empirically and is compared with various gamma sources and $\mathrm{^{12}B}\ \beta-$decay spectrum, as shown in Fig.~\ref{fig:nonL} \cite{An:2013zwz}.


\section{Inverse Beta Decay Event }
\subsection{Event Selection }
Antineutrino events are detected via the inverse beta decay (IBD) process, $\overline{\nu}_e + p \rightarrow e^+ + n$.
The positron slows down and annihilates with electron to give a prompt signal. The neutron is thermalized and then captured on Hydrogen (nH) or Gadolinium (nGd) to produce a delayed signal. The delayed signal from the nH capture emits a 2.2 MeV gamma, while the nGd capture produces several gammas with a total energy of $\sim$8 MeV.

The first step for the IBD candidate selection is to remove instrumental background caused by occasional flashing of some PMTs. Muon events are classified into three types, water pool muon ($\mu_{\mathrm{WP}}$), AD muon ($\mu_{AD}$) and shower muon ($\mu_{shower}$). The water pool muon events are identified by the number of PMT hits, NHit, in IWS or OWS. If NHits are greater than 12, these events are tagged as water pool muons. Events in AD with energies greater than 20 MeV are classified as AD muons while those with energies greater than 2.5 GeV are classified as shower muons. The non-related events occurring within the time window are rejected. The muon veto time and IBD selection criteria for both nH and nGd analysis are summarized in Table~\ref{tab:criteria}.

\begin{table}[t]
\caption[]{The IBD candidates are selected by the following criteria. For neutron captured on Gadolinium (nGd analysis), $0.7<E_p<12.0$ MeV, $6.0<E_d<12.0$ MeV, and $1<\Delta t<200 \mu s$, where $E_p$ ($E_d$) is the prompt (delayed) energy and $\Delta t=t_d-t_p$ is the time difference between the prompt and delayed signals. For neutron captured on Hydrogen (nH analysis), the prompt energy is between 1.5 and 12.0 MeV, delayed energy is within 3 $\sigma$ from the peak, and $1<\Delta t<400 \mu s$. The criteria of distance between prompt and delayed signals, $D_{pd}$, apply to nH signals.}
\label{tab:criteria}
\vspace{0.4cm}
\begin{center}
\begin{tabular}{|c|c|c|}
\hline
	& nH& nGd \\ \hline
	\mco{3}{|c|}{Reject PMT Flashers} \\ \hline
$\mu_{WP}$ veto	& 0.4 $\mu s$& 0.6 $\mu s$ \\ \hline
$\mu_{AD}$ veto	& 0.8 $\mu s$& 1 $\mu s$ \\ \hline
$\mu_{shower}$ veto	& 1 s & 1 s \\ \hline
$E_p$ [MeV] & [1.5, 12] & [0.7,12] \\ \hline
$E_d$ [MeV] & 3 $\sigma$ ($\sigma \sim 0.14\ \mathrm{MeV}$)& [6,12] \\ \hline
$\Delta t_{pd}$ [$\mu s$] & [1, 400] & [1, 200] \\ \hline
$ D_{pd}$ [mm] & 500 & N/A \\ \hline

\end{tabular}
\end{center}
\end{table}

\subsection{Background Sources }
The most important background is accidental background which are from single events and 'accidentally' pass the IBD event selection. The second effective backgrounds are from cosmic ray muons. Muon-induced products, such as fast neutron and $\mathrm{^9Li/^8He}$, can mimic IBD as a correlated pair. For the fast neutron case, neutron scattering followed by neutron capture could mimic the IBD event. For the $\mathrm{^9Li/^8He}$ background, the prompt signals are from the $\beta-$decay and the delayed signals are from neutron capture. The calibration source, AmC, in ACU is another background source. 

\section{Recent Results}

\subsection{Oscillation Parameters from Neutron Captured on Gadolinium Analysis}
With 621 days of 6-AD and 8-AD data, 150255 (613813 and 477144) antineutrino candidates with nGd capture were detected in the far site (near sites) detectors. This represents four times higher statistics than previously published results \cite{An:2013zwz}.

\begin{figure}[htbp]
\centering
\includegraphics[width=0.9\linewidth]{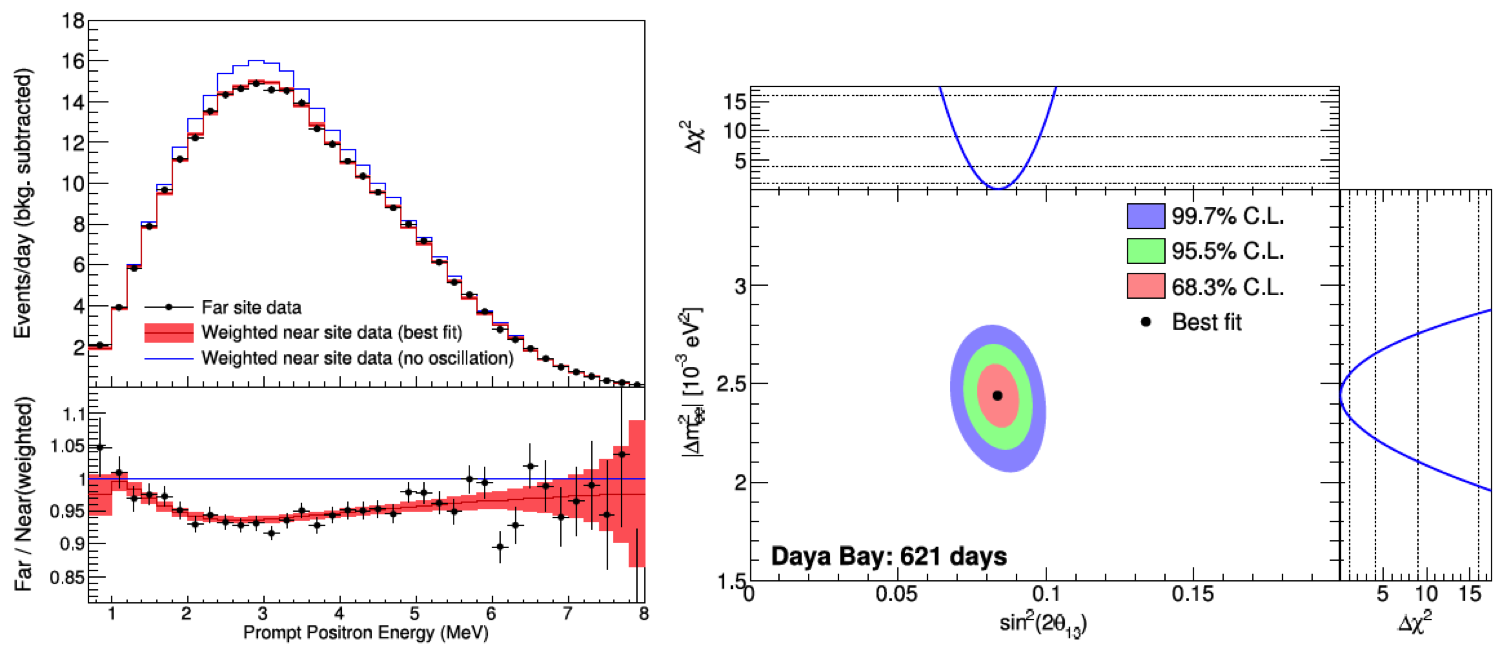}
\caption{(Preliminary) Left: the spectrum deficit in the far site. Right: The allowed regions for the neutrino oscillation parameters $\sin^2 2\theta_{13}$ and $|\Delta m^2_{ee}|$ at the 68.3, 95.5 and 99.7$\%$ confidence levels. The best estimate of the oscillation parameters are given by the black dot.}
\label{fig:results}
\end{figure}

Current preliminary results have analyzed the spectral information taking into account the nonlinearity correction and various backgrounds discussed earlier. The relative spectral distortion, as shown in Fig.~\ref{fig:results} (left), are highly consistent with oscillation interpretation. The best-fit values are $\sin^2 2 \theta_{13}=0.084 \pm 0.005$ and $|\Delta m^2_{ee}|=2.44^{+0.10}_{-0.11} \times 10^{-3} \mathrm{eV^2}$, as shown in Fig.~\ref{fig:results} (right). The precision for $\theta_{13}$ is 3 $\%$. Under the assumption of normal (inverted) neutrino mass hierarchy, the results of $|\Delta m^2_{ee}|$ is equivalent to $\Delta m^2_{32}=2.39^{+0.10}_{-0.11} \times 10^{-3} \mathrm{eV^2}$ ( $\Delta m^2_{32}=-2.49^{+0.10}_{-0.11} \times 10^{-3} \mathrm{eV^2}$ ). These results ate consistent with those from the muon neutrino disappearance experiments \cite{Adamson:2013whj, Abe:2013xua}.

\subsection{Oscillation Parameter from Neutron Captured on Hydrogen Analysis}
\begin{figure}[htbp]
\centering
\includegraphics[width=0.5\linewidth]{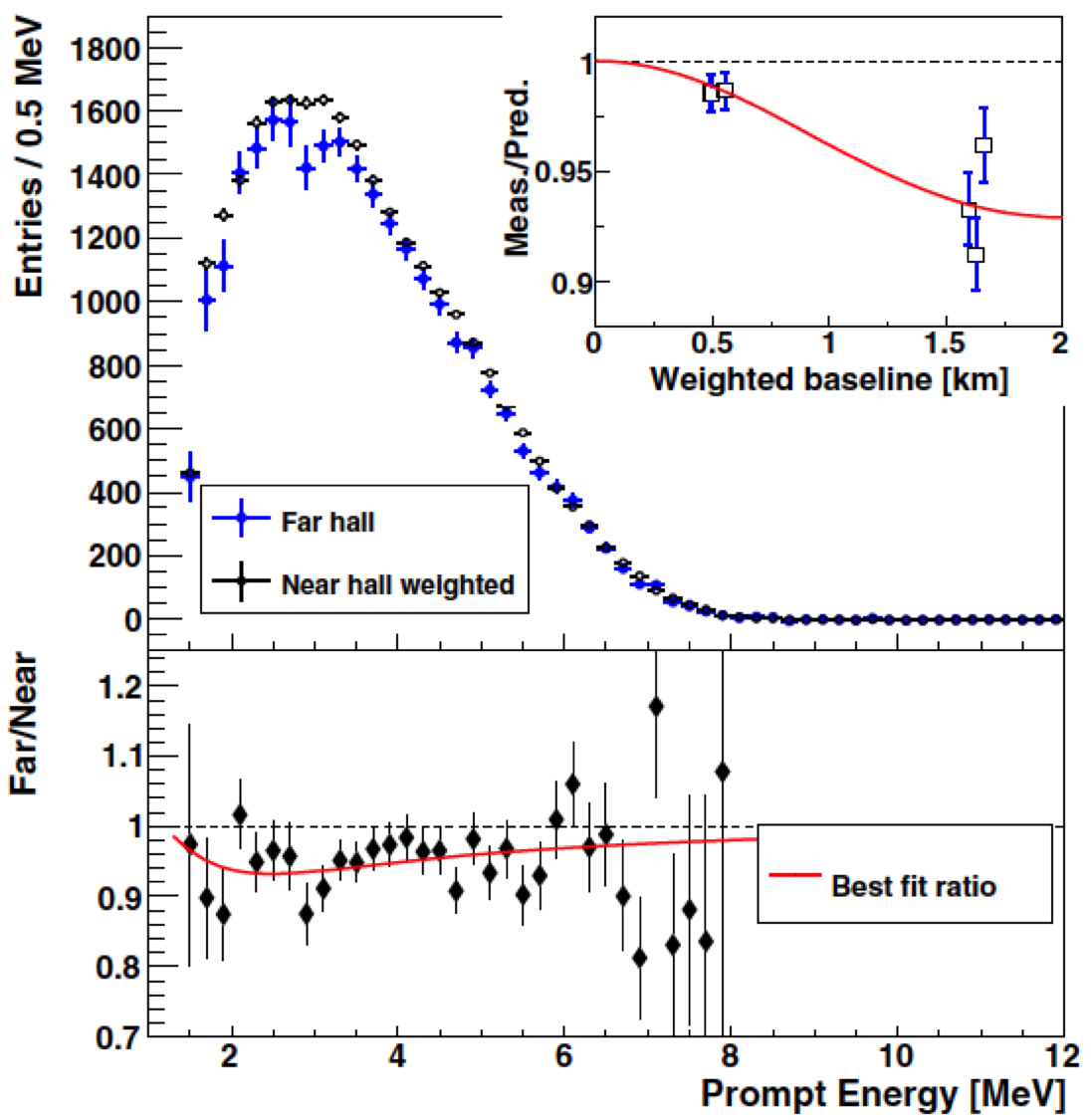}
\caption{The detected energy spectrum of the prompt events of the far-hall ADs (blue) and near-hall ADs (open circle) weighted according to baseline. The far-to-near ratio (solid dot) with the best-fit $\theta_{13}$ value is shown in the lower plot. In the inset is the ratio of the measured to the predicted rates in each AD vs baseline, in which the AD4 (AD6) baseline was shifted relative to that of AD5 by 30 (-30) m.}
\label{fig:nH}
\end{figure}

IBD events can be identified via neutrons captured on Hydrogen signals. In this study, the statistical and the major systematic uncertainties are independent from the previous Gd capture study. Several new techniques were developed to meet the challenges from the higher background and different systematics due to the lower neutron capture energy (2.2 MeV), the longer capture time (200 $\mu s$), and the larger energy loss at the detector boundary. With the 217 days of data set from the 6AD period, the rate deficit observed at the far hall is interpreted as $\sin^2 2 \theta_{13}=0.083 \pm 0.018$ with $\chi^2/NDF = 4.6/4$, as shown in Fig.~\ref{fig:nH}. The result has been combined with previous six detectors nGd analysis to give $\sin^2 2 \theta_{13}=0.089 \pm 0.008$ \cite{An:2014ehw}. Current nH analysis with the 8-AD data set is on-going.

\subsection{Sterile Neutrino Study}
\begin{figure}[htbp]
\centering
\includegraphics[width=0.5\linewidth]{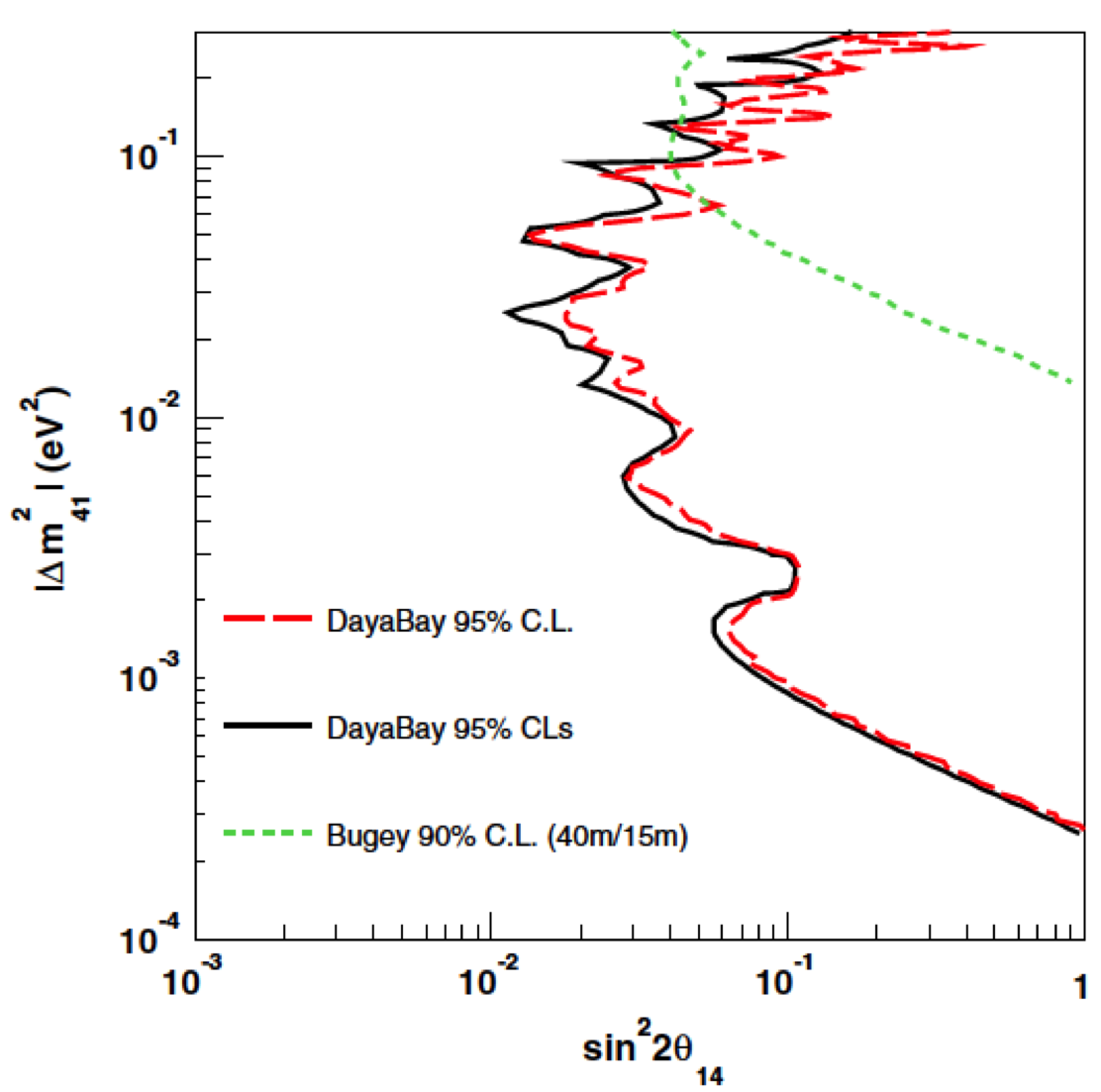}
\caption{Exclusion contours for the neutrino oscillation parameters $\sin^2 2\theta_{14}$ and $ |\Delta m^2_{41}|$. Normal mass hierarchy is assumed for both $ |\Delta m^2_{31}|$ and $ |\Delta m^2_{41}|$. The parameter space to the right side of the contours is excluded. The red long-dashed curve represents the 95$\%$ C.L. exclusion contour with Feldman-Cousins method. The black solid curve represents the 95$\%$ $\mathrm{C.L_s.}$ exclusion contour. The green dashed curve represents Bugey's 90$\%$ $\mathrm{C.L.}$ limit on $\overline{\nu}_e$ disappearance. }
\label{fig:sterile}
\end{figure}

The multiple baselines from six 2.9 $\mathrm{GW_{th}}$ nuclear reactors to six antineutrino detectors make it possible to search for light sterile neutrino in the Daya Bay experiment. With the 217 days of data set from the 6AD period, the analysis showed no evidence for sterile neutrino mixing and the most stringent limit was set at $10^{-3} \mathrm{eV^2} < |\Delta m^2_{41}| < 0.1 \mathrm{eV^2}$. Fig.~\ref{fig:sterile} shows the exclusion contours, which were determined using both the Feldman-Cousins method and the CLs method \cite{An:2014bik}.

\section{Summary }
The Daya Bay experiment has new preliminary measurements with the data set from Dec-24, 2011 to Nov-30, 2013. The results of oscillation parameters from the neutron captured on Gadolinium analysis are $\sin^2 2 \theta_{13}=0.084 \pm 0.005$ and $|\Delta m^2_{ee}|=2.44^{+0.10}_{-0.11} \times 10^{-3} \mathrm{eV^2}$. This is the most precise measurement of $\sin^2 2\theta_{13}$ to date. The independent neutron captures on Hydrogen rate analysis has measured $\sin^2 2 \theta_{13}=0.083 \pm 0.018$ with $\chi^2/NDF=4.6/4$. The sterile neutrino search has set stringent limits at $10^{-3} \mathrm{eV^2} < |\Delta m^2_{41}| < 0.1 \mathrm{eV^2}$. The data-taking for the Daya Bay experiment is planned to continue to 2017 with eight detectors. The precision of oscillation parameters, $\sin^2 2 \theta_{13}$ and $|\Delta m^2_{ee}|$, are expected to reach $3 \%$.

\section*{Acknowledgments}
Daya Bay is supported in part by the Ministry of Science and Technology of China, the U.S. Department of Energy, the Chinese Academy of Sciences, the National Natural Science Foundation of China, the Guangdong provincial government, the Shenzhen municipal government, the China General Nuclear Power Group, Key Laboratory of Particle Physics and Radiation Imaging (Tsinghua University), the Ministry of Education, Key Laboratory of Particle Physics and Particle Irradiation(Shandong University), the Ministry of Education, Shanghai Laboratory of Particle Physics and Cosmology, the Research Grants Council of the Hong Kong Special Administrative Region of China, the University Development Fund of The University of Hong Kong, the MOE program for Research of Excellence at National Taiwan University, National Chiao-Tung University, and NSC fund support from Taiwan, the U.S. National Science Foundation, the Alfred P. Sloan Foundation, the Ministry of Education, Youth, and Sports of the Czech Republic, the Joint Institute of Nuclear Research in Dubna, Russia, the CNFC-RFBR joint research program, the National Commission of Scientific and Technological Research of Chile, and the Tsinghua University Initiative Scientific Research Program. We acknowledge Yellow River Engineering Consulting Co., Ltd., and China Railway 15th Bureau Group Co., Ltd., for building the underground laboratory. We are grateful for the ongoing cooperation from the China General Nuclear Power Group and China Light and Power Company. 


\end{document}